\documentclass[letterpaper,conference,10pt]{IEEEtran}
\usepackage[letterpaper,
            left=0.625in,right=0.625in,top=0.75in,bottom=1in]{geometry}
            
\usepackage{ifluatex}
\ifluatex
  \usepackage[utf8]{luainputenc}
\else
  \usepackage[utf8]{inputenc}
\fi

\usepackage{amsmath}
\usepackage{amssymb}
\usepackage{xcolor}
\usepackage{tikz}
\usetikzlibrary{patterns}
\usepackage{pgfplots}
\usepackage{pgfplotstable}
\usepackage{adjustbox}
\usepackage{balance}

\usepackage{graphicx}
\usepackage{color}
\usepackage[caption=false, font=footnotesize]{subfig}
\usepackage{booktabs}
\usepackage{siunitx}
\usepackage{url}
\usepackage{nicefrac}
\usepackage{float}
\usepackage[giveninits=true,mincitenames=1,maxcitenames=1,maxbibnames=1,backend=biber,doi=false,isbn=false,url=false]{biblatex} %
\ifluatex
  \usepackage{fontspec}
  \setmathrm{TeX Gyre Termes}
  \setmathsf{TeX Gyre Heros}
  \setmathtt{Tex Gyre Cursor}
  \usepackage[english]{selnolig}
\fi

\usepackage{paralist}
\usepackage{tensor}
\usepackage{listings}
\input{cfdlang.sty}
\usepackage{csquotes}

\usepackage{comment}

\newboolean{showcomments}
\setboolean{showcomments}{false}
\ifthenelse{\boolean{showcomments}}
{ \newcommand{\mynote}[3]{
		\fbox{\bfseries\sffamily\scriptsize#1}
		{\small$\blacktriangleright$\textsf{\emph{\color{#3}{#2}}}$\blacktriangleleft$}}}
{ \newcommand{\mynote}[3]{}}
\newcommand{\kf}[1]{\mynote{KF}{#1}{black!20!green}}

\usepackage{cleveref}

\begin{document}
\title{From Domain-Specific Languages to Memory-Optimized Accelerators for Fluid Dynamics\looseness=-1}

\author{
\IEEEauthorblockN{Karl F. A. Friebel$^1$, Stephanie Soldavini$^2$, Gerald Hempel$^1$, Christian Pilato$^2$, Jeronimo Castrillon$^1$}
\IEEEauthorblockA{$^1$Compiler Construction Chair, cfaed, Technische Universität Dresden, Dresden, Germany\\\url{{karl.friebel, gerald.hempel, jeronimo.castrillon}@tu-dresden.de}\\
$^2$Dipartimento di Elettronica, Informazione e Bioingegneria, Politecnico di Milano, Milano, Italy\\\url{{stephanie.soldavini, christian.pilato}@polimi.it}}
}

\IEEEtitleabstractindextext{\begin{abstract}
Many applications are increasingly requiring numerical simulations for solving complex problems.
Most of these numerical algorithms are massively parallel and often implemented on parallel high-performance computers. However, classic CPU-based platforms suffer due to the demand for higher resolutions and the exponential growth of data. FPGAs offer a powerful and flexible alternative that can host accelerators to complement such platforms. Developing such application-specific accelerators is still challenging because it is hard to provide efficient code for hardware synthesis. 
In this paper, we study the challenges of porting a numerical simulation kernel onto FPGA.
We propose an automated tool flow from a domain-specific language (DSL) to generate accelerators for computational fluid dynamics on FPGA. 
Our DSL-based flow simplifies the exploration of parameters and constraints such as on-chip memory usage.
We also propose a decoupled optimization of memory and logic resources, which allows us to better use the limited FPGA resources.
In our preliminary evaluation, this enabled doubling the number of parallel kernels, increasing the accelerator speedup versus ARM execution from 7 to 12 times.
\end{abstract}%
}

\maketitle
\IEEEdisplaynontitleabstractindextext
\IEEEpeerreviewmaketitle
\section{Introduction}
\label{sec:intro}

In the last years, data-intensive applications have permeated many computing areas due to the surge of deep learning and the ever-increasing demand for resolution in physics simulations (e.g., molecular dynamics, weather simulations).
At the same time, the diminishing returns of technology scaling has led to vast system heterogeneity, with GPUs and tensor accelerators~\cite{chen2014diannao,jouppi2017datacenter, 9286149}. 
Hardware accelerators achieve high performance and energy efficiency thanks to specialization and spatial parallelism~\cite{10.1145/3361682}. 
Reconfigurable hardware, like FPGA devices, is an attractive solution to democratize the use of such accelerators for different users~\cite{pilato_date21}.

Despite the progress in high-level synthesis (HLS)~\cite{7368920}, we still face a large \emph{semantic gap} between application experts and FPGA hardware architects for FPGA-based systems. 
This paper targets physicists and numerical experts for computational fluid dynamics (CFD). In this domain, the researchers must not only adapt the algorithms for a particular simulation but also face fragmented tools, integration tasks, complex libraries, and HLS directives for different targets.
Integration tools can tackle such complexity by raising the abstraction level with language support, or by improving analysis and optimization to generate hardware from low-level code like C or Fortran.

Research on hardware generation from low-level code has a long history, with methods dating back to early auto-parallelising compilers~\cite{girkar1992automatic}.
Recent advances in code analysis and optimization reduce the manual effort required to produce HLS-friendly code, for instance, by inserting HLS pragmas~\cite{santos2020automatic} or by generating optimized systolic arrays~\cite{autosa}.
Parallelism extraction is however sensitive to coding style.  
An alternative is to express a high-level specification with rich semantics in the form of a domain-specific language (DSL). 
High-performance DSLs have been successfully used to target CPUs and GPUs, e.g., for image processing~\cite{pub:halide}, general tensor computations~\cite{kjolstad2017tensor, rink_gpce18}, and deep learning~\cite{chen2018tvm, tensor_comprehensions}. 
Similar flows have been proposed for FPGAs~\cite{spatial,8735529}.
Since most DSLs have high-level operators and data structures, compilers can decide shapes, layouts, and \emph{schedules} to generate target-aware code.

In this paper we present a proof-of-concept of an end-to-end methodology that leverages the high-level semantics in DSLs to create FPGA-based systems and accelerate numerical kernels in CFD simulations (cf. Section~\ref{sec:approach}). 
As in other DSLs, we lower our specification into a polyhedral model, allowing us to leverage existing polyhedral transformations. 
Our key contributions are: 
(1) a study of code generation strategies to produce code that is amenable to commercial HLS (cf. Section~\ref{sec:compiler}), and 
(2) an approach to decouple the computational logic from management of on-chip data to improve the overall system efficiency and create composable architectures (cf. Section~\ref{sec:hardware}). 
Decoupling computation from data management is particularly important for CFD and data-intensive applications to better coordinate data exchanges with the host CPU, hide/reduce the communication latency, and increase parallelism. 
Our decoupled approach is relevant also for other DSL-to-hardware compilation flows. 
We provide a preliminary evaluation for 
a fundamental CFD kernel, which helps identifying the potential and the challenges for upcoming FPGA nodes in HPC (cf. Section~\ref{sec:eval}). 
For example, by exploiting memory sharing, we can fit more parallel kernel instances, increasing the speedup from $7.09\times$ to $12.58\times$ compared to ARM execution.

%\subsection{A DSL for Spectral Element Methods}
\section{Background on Fluid Dynamics}\label{sec:background}
\subsection{Spectral Element Methods}
\label{sec:sem}

In numerical mathematics, spectral element methods~(SEM) are common in solving partial differential equations (PDEs), like the Navier-Stokes equations, which are impossible to solve analytically. 
SEM approximates the solution using functions, like the Fourier series, transforming the unknown physical quantities of the problem into spectral coefficients.

To reduce the numerical complexity, the simulated volume is divided into~$N_eq$ smaller volumes.
By partitioning the total space into several sub-spaces or \textit{elements}, SEM reduces the numerical error introduced by the approximation.
To further reduce the error, SEM uses an approximation based on polynomials of a higher degree ($p>1$).
The solution is expressed as a linear system of equations which can be solved locally for each element.
An element solution~$e$ can be represented in three dimensions as a tensor~$v_ {ijk, e}$ with~$i,j,k \in \{0, \dots, p\}$. Often, the polynomial degree~$p$ is the same for all spatial dimensions.

In this paper we focus on the Helmholtz equations, which are common in PDE solvers. 
Moreover, the Inverse Helmholtz operator is complex enough to subsume simpler operators (e.g., interpolation) which are similarly relevant in CFD simulations~\cite{huismann_2017}. 
The operator can be formulated as: 
\begin{subequations}
  \begin{align}
    \label{eq:helm_op:1}
    t_{ijk,e} &= \sum_{l=0}^{p} \sum_{m=0}^{p} \sum_{n=0}^{p} S^T_{li} \cdot S^T_{mj} \cdot S^T_{nk} \cdot u_{lmn,e} \\
    \label{eq:helm_op:2}
    r_{ijk,e} &= D_{ijk,e} \cdot t_{ijk,e}\\
    \label{eq:helm_op:3}
    v_{ijk,e} &= \sum_{l=0}^{p} \sum_{m=0}^{p} \sum_{n=0}^{p} S_{li} \cdot S_{mj} \cdot S_{nk} \cdot r_{lmn,e} \, .
  \end{align}
\end{subequations}

\subsection{CFDlang DSL}
\label{sec:CFDlang}
In this paper we extend the CFDlang DSL for tensor operations~\cite{cfdlang}. 
CFDlang is target-agnostic and offers the user an interface that is close to the mathematical problem specification.
The CFDlang notation is motivated by the tensor product notation often found in CFD applications.
Equations~\ref{eq:helm_op_1:1}-\ref{eq:helm_op_1:3} are all equivalent to Equation~\ref{eq:helm_op:1}.

\begin{figure}
\begin{lstlisting}[language=CFDlang,numbers=left,numbersep=5pt,basicstyle=\footnotesize\ttfamily,frame=tb,numberstyle=\tiny,linewidth=0.95\columnwidth,xleftmargin=2em,framexleftmargin=1.5em]
var input S   : [11 11]
var input D   : [11 11 11]
var input u   : [11 11 11]
var output v  : [11 11 11]
var t         : [11 11 11]
var r         : [11 11 11]
t = S # S # S # u . [[1 6] [3 7] [5 8]]
r = D * t
v = S # S # S # r . [[0 6] [2 7] [4 8]]
\end{lstlisting}
\vspace{-2mm}%
\caption{DSL code for the Inverse Helmholtz operator.}%
\label{fig:helm_op:dsl}%
\vspace{-2mm}%
\end{figure}%

\begin{subequations}
\label{eq:helm_op_1}
  \begin{align}
    \label{eq:helm_op_1:1}
    \mathbf{v} &= \left ( \mathbf{S}^T \otimes \mathbf{S}^T \otimes \mathbf{S}^T \right ) \mathbf{u}\\
    \label{eq:helm_op_1:2}
    &= \left ( \mathbf{S}^T \otimes \mathbf{S}^T \otimes \mathbf{S}^T \otimes \mathbf{u} \right )\indices{^{iljmkn}_{lmn}}\\
    \label{eq:helm_op_1:3}
    &= \left ( \mathbf{S} \otimes \mathbf{S} \otimes \mathbf{S} \otimes \mathbf{u} \right )\indices{^{limjnk}_{lmn}}
  \end{align}
\end{subequations}

In CFDlang, Equation~\ref{eq:helm_op_1:3} could be represented as $S \, \#\, S \, \#\, S \, \#\, u \, . \, [[1\, 6] \, [3\, 7] \, [5\, 8]]$.
Here $S \, \# \, S \, \# \, S \, \# \, u$ is the outer product of all tensors involved in the contraction.
The dimensions of this product tensor are numbered from 0-8.
The index pairs in the square brackets then specify which dimensions are reduced in the contraction.
In addition CFDlang supports most of the tensor operations typically used for CFD simulations such as \emph{tensor contractions} (cf.~\ref{eq:helm_op:1} and~\ref{eq:helm_op:3}), \emph{inner} and  \emph{outer products}, and \emph{entry-wise multiplication} (cf.~\ref{eq:helm_op:2}).

Figure~\ref{fig:helm_op:dsl} shows a description of the complete Inverse Helmholtz operator in CFDlang.
Lines 1-6 describe all required tensors including the intermediate values for $t$ and $r$.
Lines 7 and 9 show a tensor contraction and line 8 contains a Hadamard product.
In CFDlang the program does not determine the order of operations and the exact implementation, allowing the compiler to optimize the operations for a particular target.

\section{System Architecture and Methodology Overview}
\label{sec:approach}

\subsection{System-level FPGA-based Design for CFD Simulations}\label{sec:target}

\begin{figure}
\centering
\includegraphics[width=0.9\columnwidth]{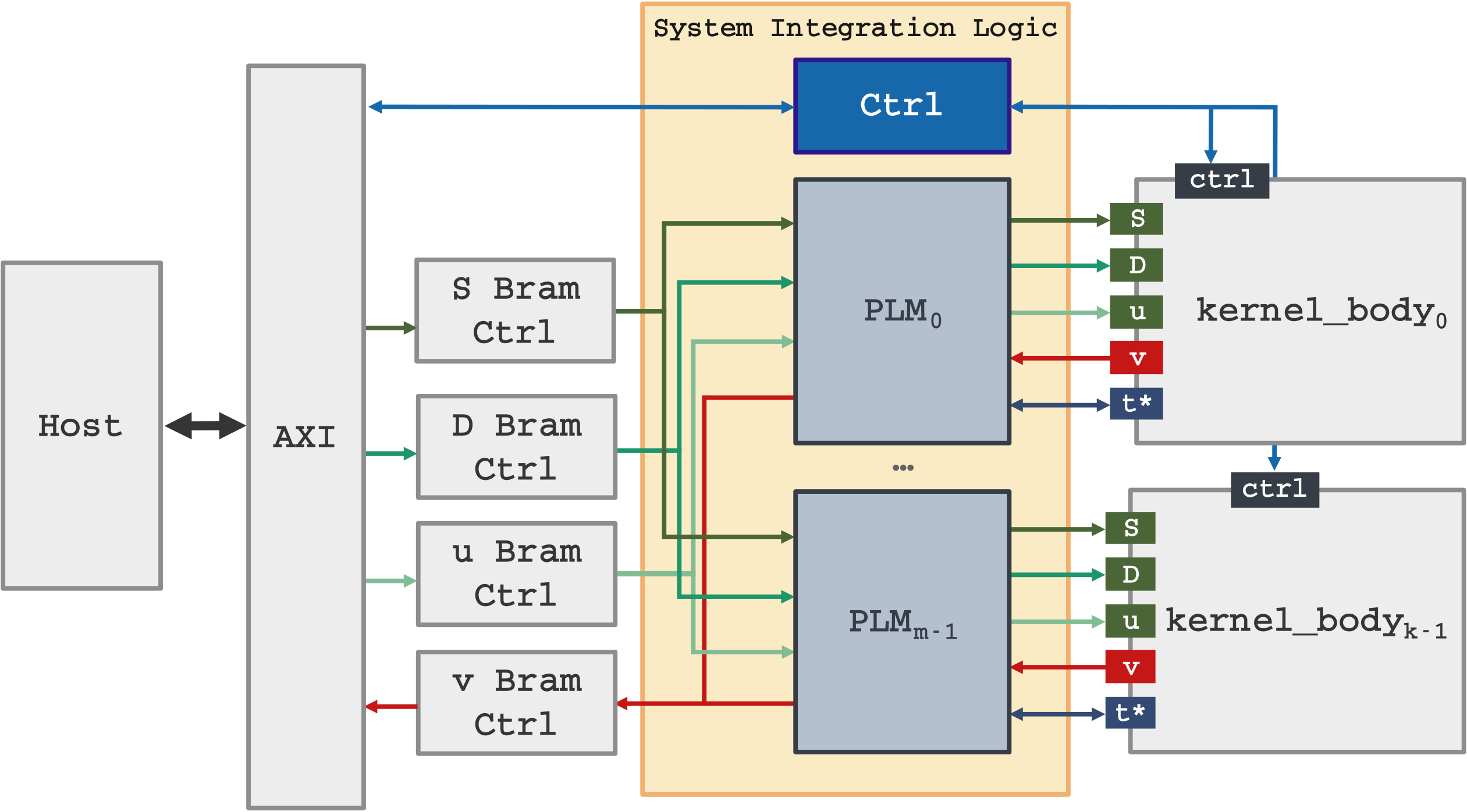}
\caption{Target system instance for the Inverse Helmholtz: we replicate the accelerator (along with its PLM) operator multiple times for parallel execution.}\label{fig:template}	
\vspace{-0.2cm}
\end{figure}

From the system-level perspective, the CFD simulation runs on the host, which sends the kernel data to the FPGA (\texttt{S}, \texttt{D}, and \texttt{u}) and retrieves the output (\texttt{v}) after the kernel execution. 
Since CFD simulations are massively parallel, we can parallelize multiple elements. 
We design each accelerator by combining HLS and Private Local Memory (PLM) optimization tools. 
This allows us to optimize the two parts independently and replicate them based on the amount of FPGA resources requested by HLS. 
\Cref{fig:template} shows an example where $m=k$ and so each PLM instance (for one element) is associated with the corresponding kernel. If $k < m$ (e.g., $m=16$ and $k=4$), the same accelerator operates on consecutive PLM elements.

\subsection{Decoupled CFDlang-to-Bitstream Flow}

We propose a modular tool flow that simplifies the creation of FPGA accelerators for numerical simulations directly from CFDlang. 
Concretely, it helps the user optimize intra-kernel and inter-kernel parallelism, and host-accelerator interfacing. 

\begin{figure}
    \includegraphics[width=\linewidth]{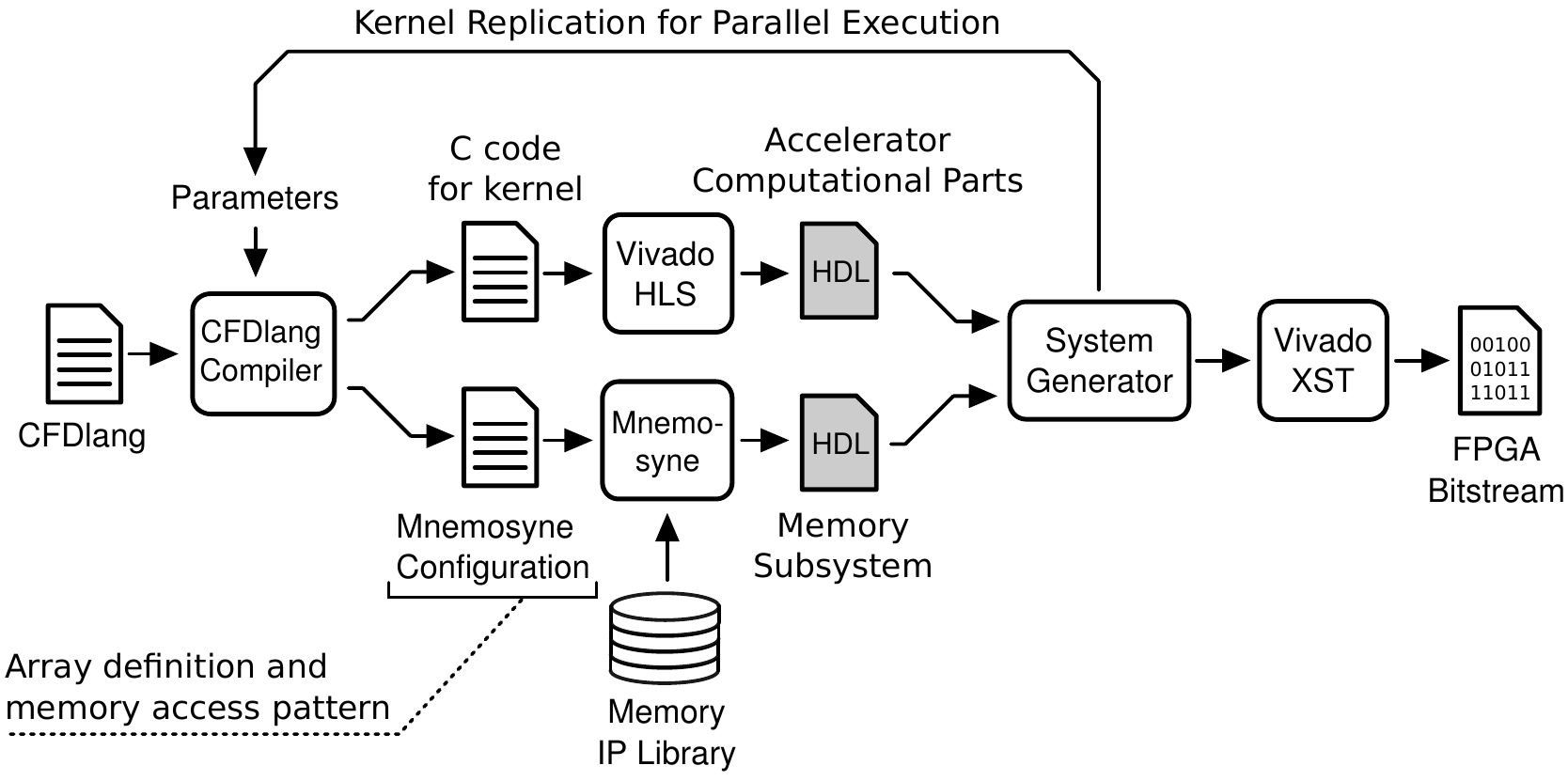}
    \caption{Tool flow from CFDlang to FPGA bitstream generation.}\label{fig:toolflow}
\end{figure}

Figure~\ref{fig:toolflow} shows an overview of our flow. 
The CFDlang's representation gives us fine-grained control to rearrange data accesses or modify the number of parallel accesses.
In this work we extend the CFDlang compiler infrastructure with FPGA-specific optimizations and hardware generation. 
We added a polyhedral engine, using libISL~\cite{pub:isl}, for intra-kernel transformations for HLS and memory optimization. 

The data layout and the kernel implementation generated by the compiler are then optimized separately. 
We use commercial HLS (currently Vivado HLS) to generate the accelerator implementation, from C code generated from the compiler. 
We use high-level operator information in the DSL and leverage polyhedral analysis to fine-tune the generated code so that it is amenable for HSL (cf. Section~\ref{sec:compiler}). 
This source-to-source approach allows us to profit from excellent results from HLS tools for the computational part.  
Classic HLS tools, however, have limited support for implementing  multi-bank, multi-port memories. 
For this reason we use Mnemosyne~\cite{7572091}, which takes over the generation of the memory architecture for the accelerator and supports us in the effective use of FPGA BRAMs. 
We modified the CFDlang compiler to automatically create the Mnemosyne input metadata during the compilation (cf. Section~\ref{sec:liveness}). 
This is crucial since the compiler can support sophisticated partitioning or sharing of data among multiple memory banks through code analysis. 

The system generator in Figure~\ref{fig:toolflow} automatically creates the logic for replicating the kernels (produced by HLS) and the memories (produced by Mnemosyne). 
The tool flow finally produces the artifacts for interfacing with bitstream generation and the corresponding host software to control the accelerators.

From the perspective of an application developer, we enable a seamless integration of the CFDlang in Fortran or C++ code. 
The kernel with the respective accelerator is then called via a predefined function handle from the surrounding application.

\section{DSL Lowering}
\label{sec:compiler}

\subsection{CFDlang Compiler Extension: Overview}
\label{sec:CFDlang-ext}
\begin{figure}
    \centering
    \includegraphics[scale=0.54]{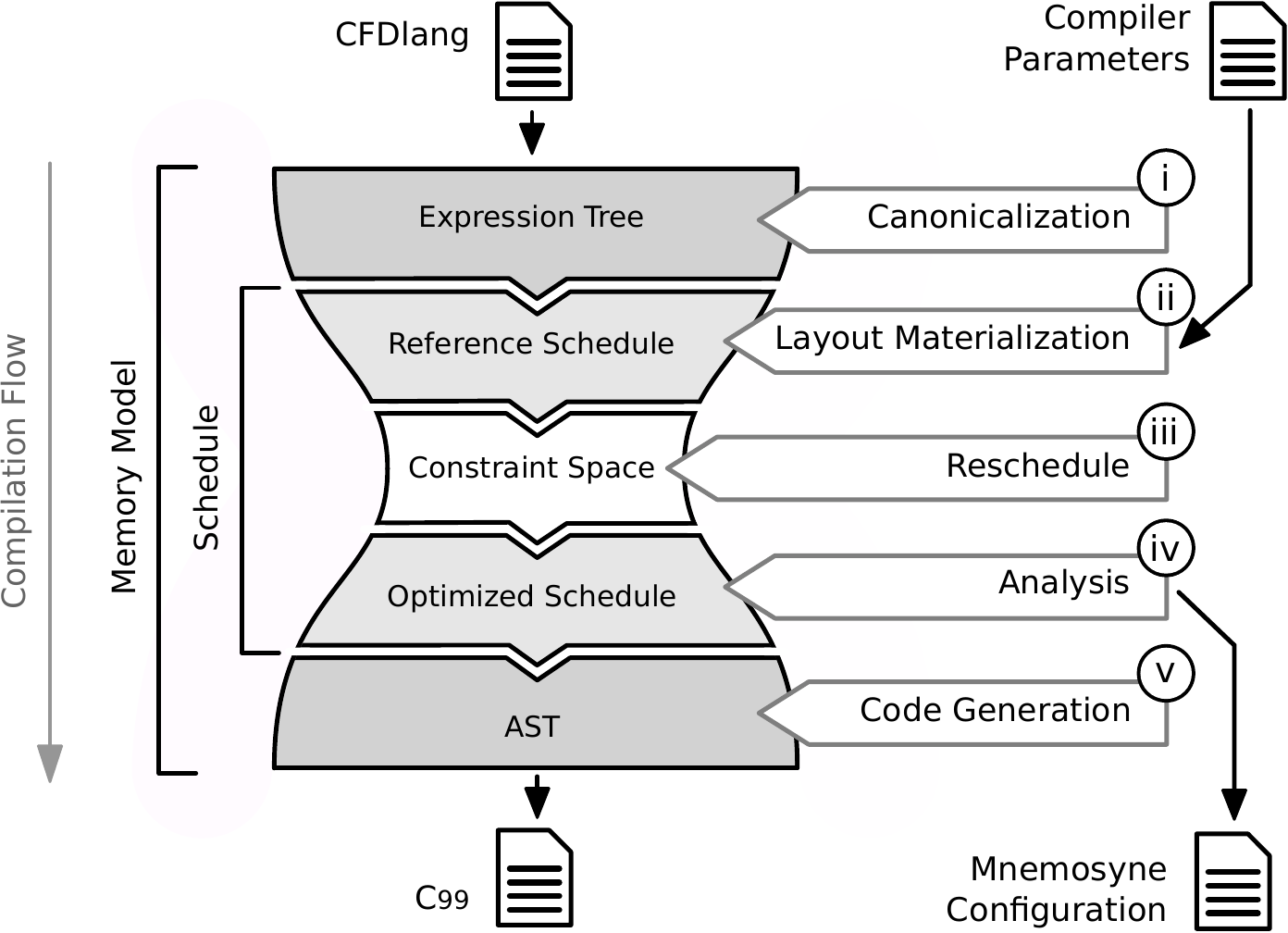}
    \caption{The extended CFDlang compilation flow. This diagram shows the different levels of abstraction and the operations performed on them.}\label{fig:CFDlang}
\end{figure}

Figure~\ref{fig:CFDlang} shows an overview of our current CFDlang compiler. 
As discussed in Section~\ref{sec:background}, the compiler accepts tensor programs such as the Inverse Helmholtz kernel (cf. Figure~\ref{fig:helm_op:dsl}). 

The CFDlang frontend creates a simple intermediate representation (IR) that models each statement by constructing an expression tree for the right-hand side (RHS).
In this representation, the compiler can detect the independence of reduction dimensions in contraction expressions to exploit associativity. 
This allows transforming \Cref{eq:helm_op_1:3} into an equivalent expression that computes multiple reductions of lower ranks:
\begin{equation*}
    \mathbf{t} = \left (\mathbf{S} \otimes \left (\mathbf{S} \otimes \left (\mathbf{S} \otimes \mathbf{u} \right )\indices{^{cz}_{xyz}} \right )\indices{^{by}_{cxy}} \right )\indices{^{ax}_{bcx}}
\end{equation*}
These transformations operate entirely on the IR and are the basis for existing CFDlang optimizations.

As shown in Figure~\ref{fig:CFDlang}, we extended the compiler by gradual abstraction, incrementally lowering the input to a more flexible and concise constraint-based description.
We apply existing transforms during step \textcircled{\small i} before introducing any new abstractions.
Our flow can further specialize this abstract representation to achieve more HLS-friendly kernels.
The process stops once a rigid C99 implementation has been reached.

\newcommand{\romancircled}[1]{{\large\textcircled{\small\Rn{#1}}}}

\subsection{Modelling Tensor Values}
\label{sec:tensorvals}

After having produced a pseudo-SSA
form in step \romancircled{1}, all expressions are fixed.
The order in which individual elements of these expressions are computed is still undefined, and the IR does not reflect the per-element dependencies.
Therefore, we introduce a value-based abstraction of tensor expressions, which differs from memory-based methods, such as the typical \texttt{memref}-based usage of the \texttt{linalg} dialect in MLIR~\cite{lattner2021mlir}. % JC: Very good!

As the language only knows statically shaped non-aliasing tensor values, we can reference any particular element of any tensor using an index tuple.
Since we build on \textit{isl}, we use its notation.
For example, the set of index tuples for the tensor $\mathbf{t}$ in the kernel from \Cref{fig:helm_op:dsl} is written as:
\begin{equation*}
    \{ \mathbf{t}[\underbrace{i,j,k}_{\text{index tuple}}] : 0 \leq i < 11 \:\text{and}\: 0 \leq j < 11 \:\text{and}\: 0 \leq k < 11 \}
\end{equation*}
Every tensor spans its own, unique index space with its rank setting the number of dimensions.
This also applies to scalars, which are modelled as 0-dimensional, and thus have exactly one valid \textquote{index} each.
When reasoning about types, we refer to any tensor index space using the shorthand $\textit{tensor}[\ldots]$.

Every expression in the IR defines all elements of a unique tensor via the RHS expression.
There are named tensors that appear on the left-hand side of an assignment, which may either be part of the kernel interface (cf. \lstinline[language=CFDlang]{input} and \lstinline[language=CFDlang]{output} in Figure~\ref{fig:helm_op:dsl}) or local temporaries like $\mathbf{t}$.
All other expressions define transient (a.k.a. virtual) tensors without an explicit name.

We can examine assignments via mappings of data dependencies $\textit{tensor}[\ldots] \rightarrow \bigcup \textit{tensor}[\ldots]$ from output to operand tensor elements. 
For the Hadamard product in Line 8, 
    $\mathbf{r} = \mathbf{D} \circ \mathbf{t}$,  
we obtain elements mappings from $\mathbf{r}$ to $\mathbf{D}$ and $\mathbf{t}$:
\begin{align*}
    \mathbf{r}[i,j,k] &\mapsto \mathbf{D}[i,j,k] \cup \mathbf{t}[i,j,k]
\end{align*}
We compute this mapping, called the \emph{operand map}, for every tensor expression by transitive application.

Consider the contraction expression on Line 7, the Equation~\ref{eq:helm_op_1:3}.
From the internal structure of the reduction, we can define an inner domain for the expression that includes the reduction indices 
$\{\mathbf{t}[i,j,k,\alpha,\beta,\gamma] : \ldots\}$, 
from which we then construct an inner operand map:
\begin{align*}
    \mathbf{t}[i,j,k,\alpha,\beta,\gamma] &\mapsto \mathbf{S}[i,\alpha] \cup \mathbf{S}[j,\beta] \cup \mathbf{S}[k,\gamma] \cup \mathbf{u}[\alpha,\beta,\gamma]
\end{align*}
To obtain the composable mapping over the outer, output tensor domain, we project out these indices.

\subsection{Computing a Reference Schedule}

In a polyhedral model, a statement $\textit{stmt}[\ldots]$ is a space over some integer control variables, and its points are called \textit{instances}.
A schedule $S : \textit{stmt}[\ldots] \rightarrow [\ldots]$ maps these instances to the schedule space, which is an anonymous integer tuple space that reflects an executable loop program structure.
These tuples impose a total ordering via lexicographical comparison, enabling a mathematical abstraction for code transformations. 

We promote every assignment to a statement, allowing us to leverage 
existing schedule optimizations for tensor expressions.
The order of the domain elements is not fixed by the CFDlang program, but there is an implicit reference schedule that defines what orders are valid.
We construct the reference schedule from the assignments and their operand maps to enable layout-aware transformations.
During construction, we use the operand maps to avoid materializing transients, and inner domain maps to lower reductions into schedule space.

\subsection{Layout Materialization}\label{sec:layout}

In step \romancircled{2} of Figure~\ref{fig:CFDlang}, we use the reference schedule to concretize tensor memory layouts as pre-optimization.
This allows us to adapt to external constraints, such as the host memory layout, and to make use of array partitions during rescheduling.
This differs from typical polyhedral approaches that perform the layout independently from their scheduling.
Instead, we use a model-driven construction of the layouts through command-line options, and modify our schedule accordingly.
Such options include \textit{layout expressions} which map tensors to arrays.
An array is a one-dimensional index space $\textit{array}[i]$, later implemented using concrete platform memory.
For example, the C99 standard innermost dimension layout of $\mathbf{t}$ reads 
    $\mathbf{t}[i,j,k] \mapsto \texttt{t}[121i + 11j + k]$.
Every tensor must have an affine layout, and we default to the row major layout.
These expressions can also be used to implement implicit reshaping as is commonly done in host-device interfaces.

% Partitioning maps.
Options include also \textit{partitioning maps} which map arrays to arrays.
These mappings can declare relations of the very general type $\bigcup \textit{array}[i] \rightarrow \bigcup \textit{array}[o]$, provided that their union has an injective fixpoint.
This means that they can, in fact, split \emph{and} merge arrays, despite the name.
This allows non-surjective mappings, which can be used to implement explicit address-space sharing if the transformation is legal (cf. Section~\ref{sec:mnemosyne}).

% Role of the reference schedule.
The reference schedule reflects these mappings after their application by transforming the statement data dependencies, and splitting the statements to operate over disjoint sets of array partitions.
As a result, the subsequent rescheduling process can independently schedule computations in different array partitions regardless of the original expression structure.

\subsection{Rescheduling and Code Generation}

% Summary.
In step \romancircled{3}, we use isl's Pluto scheduler to compute schedules from the constraints derived from the reference schedule.
We obtain these constraints through layout-aware dataflow analysis. 
We use read-after-write (RAW) dependencies as cost function in the isl rescheduler to reduce the dependence distance and thus the live intervals.
Read-after-read (RAR) dependencies also feed a cost function that attempts to place the statements at coincident schedule space points. 
This helps reduce the pressure on temporary storage.

% Code generation.
Finally, step \romancircled{5} calls isl's code generator to produce a C99 program that implements the computed schedule.
Our precomputed operand maps simplify the generation of the expressions for each element.

\subsection{Liveness Analysis}
\label{sec:liveness}

% Motivation.
To optimize the memory architecture, Mnemosyne needs external information on the memory interface.
Based on this, it applies sharing transformations based on a memory compatibility graph, which we can easily compute from the CFDlang program for any given schedule.
We perform these analyses and generate such metadata in step \romancircled{4}.

\begin{figure}
    \centering
    \vspace{-1.55em}
    \includegraphics[scale=0.41]{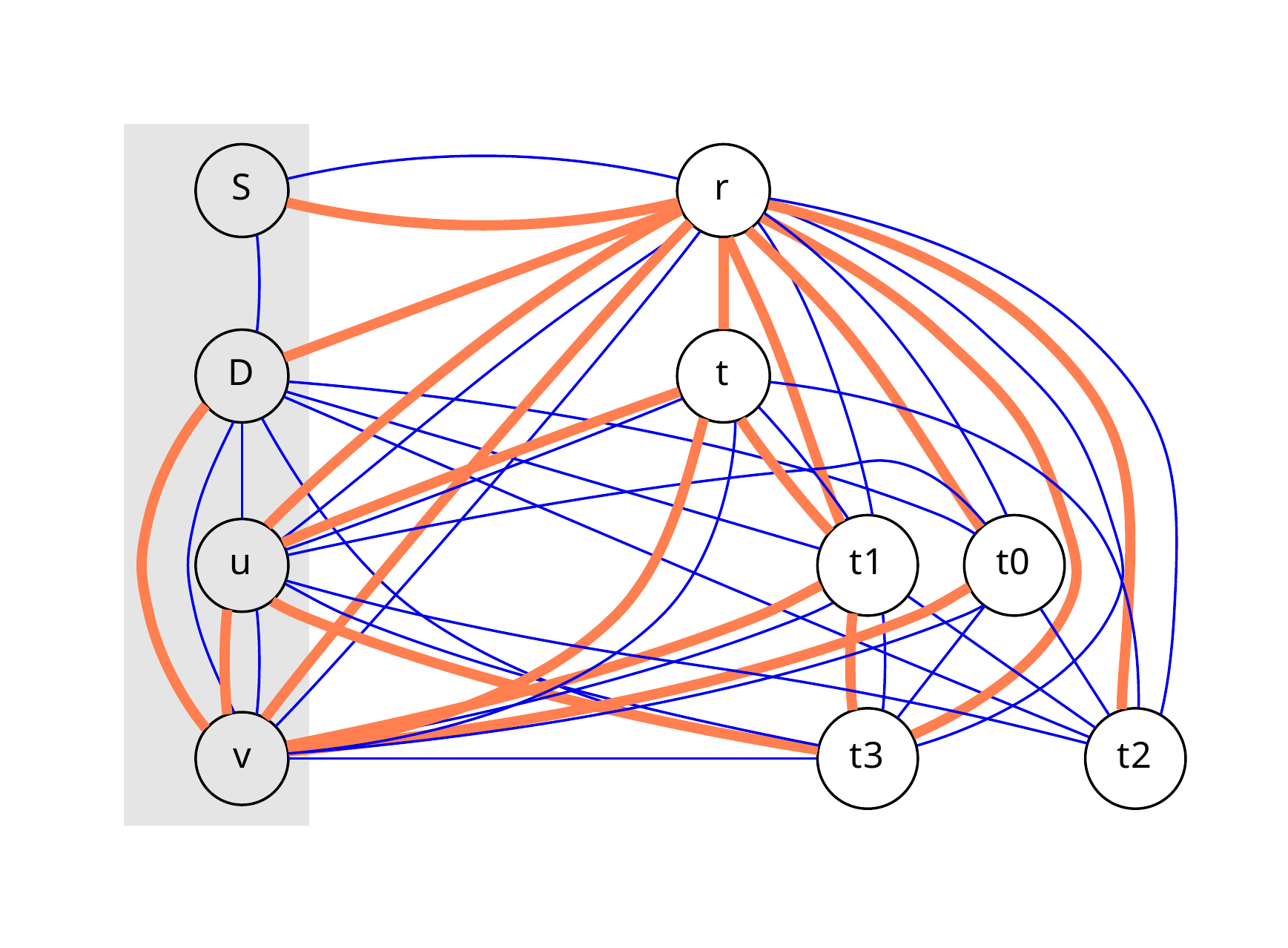}
    \vspace{-2.55em}
    \caption{Graph of the \textcolor{blue}{memory-interface} and \textcolor{orange!70!gray}{\textbf{address-space}} compatibilities for Inverse Helmholtz kernel. Interface arrays are grouped on the left.}\label{fig:mem_compat}
    \vspace{-0.3mm}
\end{figure}

% Memory compatibilities.
\Cref{fig:mem_compat} shows a memory compatibility graph derived from a valid schedule of the kernel in \Cref{fig:helm_op:dsl}.
In this graph, nodes represent arrays, with the edges indicating sharing potential.
Two arrays are \textcolor{orange!70!gray}{\textbf{address-space}} compatible if their lifetimes do not overlap for the entire execution of the accelerator.
Two arrays are \textcolor{blue}{memory-interface} compatible if it is possible to define a total temporal ordering of the memory operations such that the same type (either read or write) never happens at the same time on both arrays.

% Liveness analysis in the polyhedral model.
Dataflow analysis returns RAW dependencies in the form:
\begin{equation*}
    \texttt{RAW} : \textit{array}[i] \rightarrow [\textit{write}[\ldots] \rightarrow \text{read}[\ldots]]
\end{equation*}
This relation maps $\textit{array}$ elements that transport a value from the statement $\textit{write}$ to the statement $\textit{read}$.
By definition, the value at the specified array element is live at all schedule space points between these statements.
By applying the schedule $S$ to both statements, we transform the RAW dependencies into the liveness intervals $I = (S \times S) \circ \texttt{RAW}$ over schedule space tuples:
\begin{align*}
    I &: \textit{array}[i] \rightarrow [[\ldots] \rightarrow [\ldots]]
\end{align*}
Correctly inferring the liveness of \lstinline[language=CFDlang]{input} and \lstinline[language=CFDlang]{output} arrays requires a modified virtual schedule.
In this schedule, two statements \texttt{first} and \texttt{last} are defined, modelling writes to inputs and reads from outputs.
    
Since the schedule space tuples are lexicographically ordered, we can define a second-order helper function $\texttt{ge\_le}$ that turns a mapping from one tuple to another into a set of all tuples between them.
Finally, we obtain $L = \texttt{ge\_le} \circ I$, mapping every array element to the set of schedule tuples at which it carries a live value.
\begin{align*}
    \texttt{ge\_le} &: [[\ldots] \rightarrow [\ldots]] \rightarrow [\ldots]\\
    L &: \textit{array}[i] \rightarrow [\ldots]
\end{align*}
To determine whether two arrays are address-space compatible, one must now simply determine whether their images in $L$ are disjoint.

\begin{figure}
\includegraphics[width=\columnwidth]{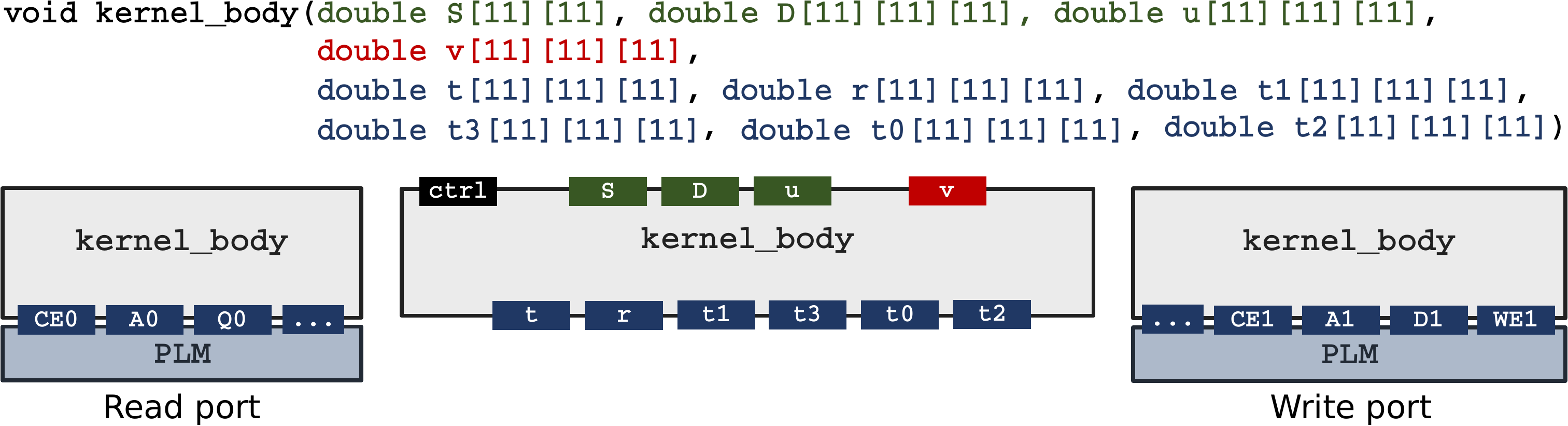}
\vspace{-1.75em}
\caption{Generation of accelerator kernel from C code: optimized PLM units store the arrays and are accessed with standard memory ports.
For readability, the interface uses multi-dimensional arrays instead of flattened 1-D arrays.
% \kf{I'd really like for us to change this figure and use the new prototype, which has just the layouted 1D arrays, even if the two are equivalent in this case.}
}\label{fig:interface}
\end{figure}

\section{Generation of Hardware Architecture}
\label{sec:hardware}

\begin{figure*}
     \centering
        \subfloat[$m=k=1$\label{fig:k1m1}]{%
           \includegraphics[scale=0.38]{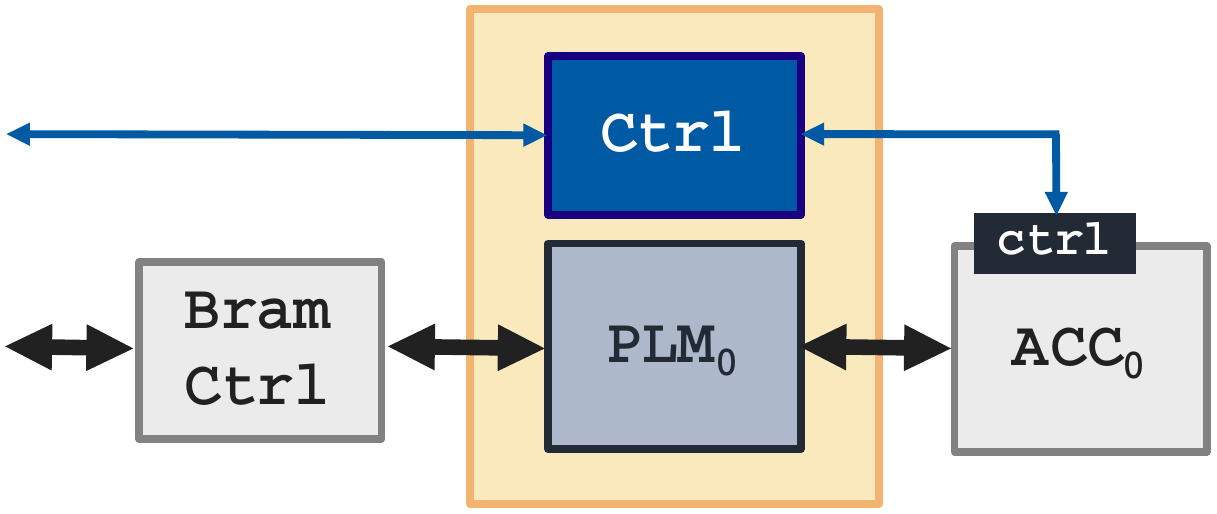}
        }
     \hfill
        \subfloat[$m=k>1$\label{fig:k2m2}]{%
           \includegraphics[scale=0.38]{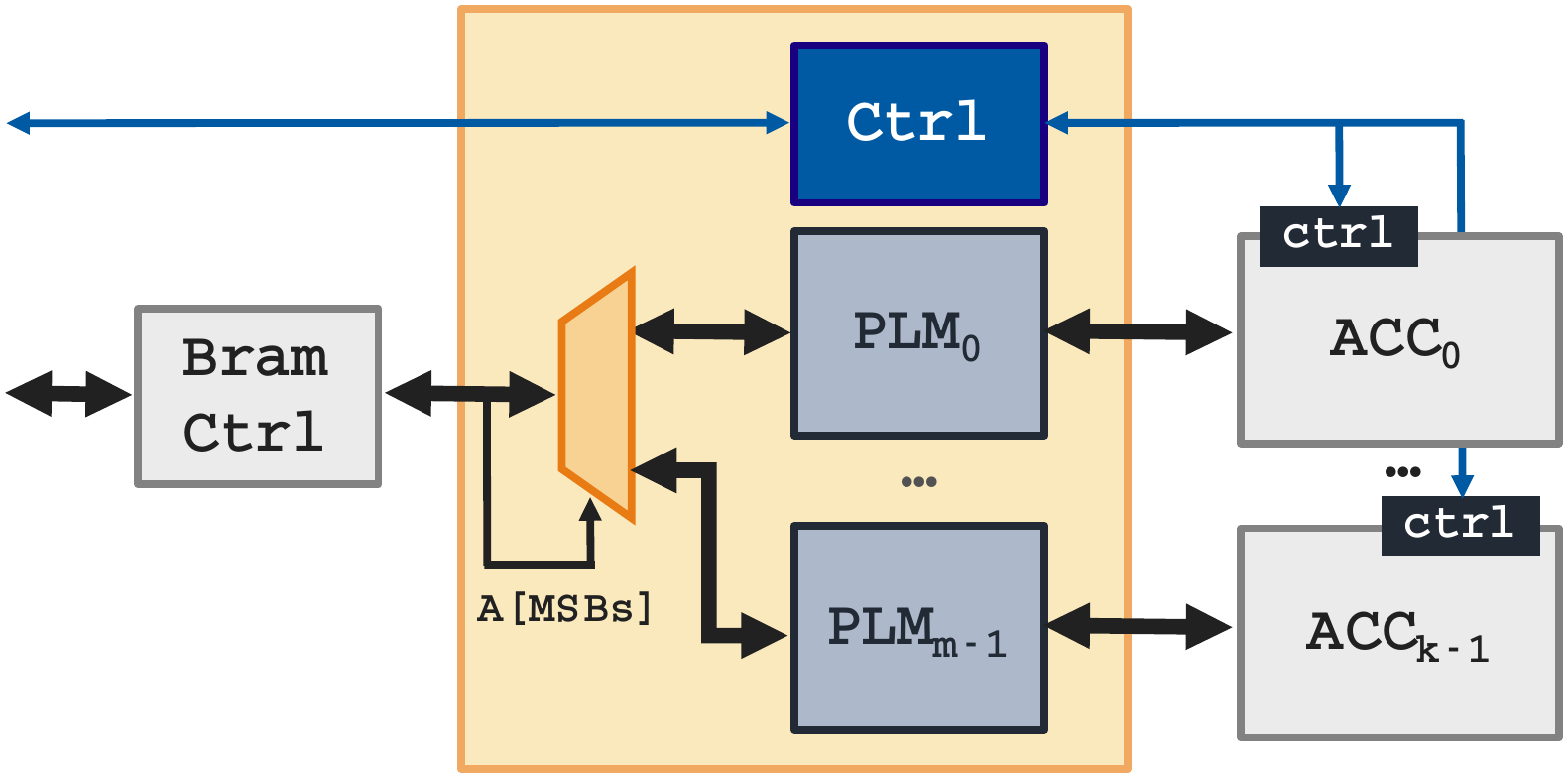}
        }
     \hfill
        \subfloat[$m>k$\label{fig:k1m2}]{%
           \includegraphics[scale=0.38]{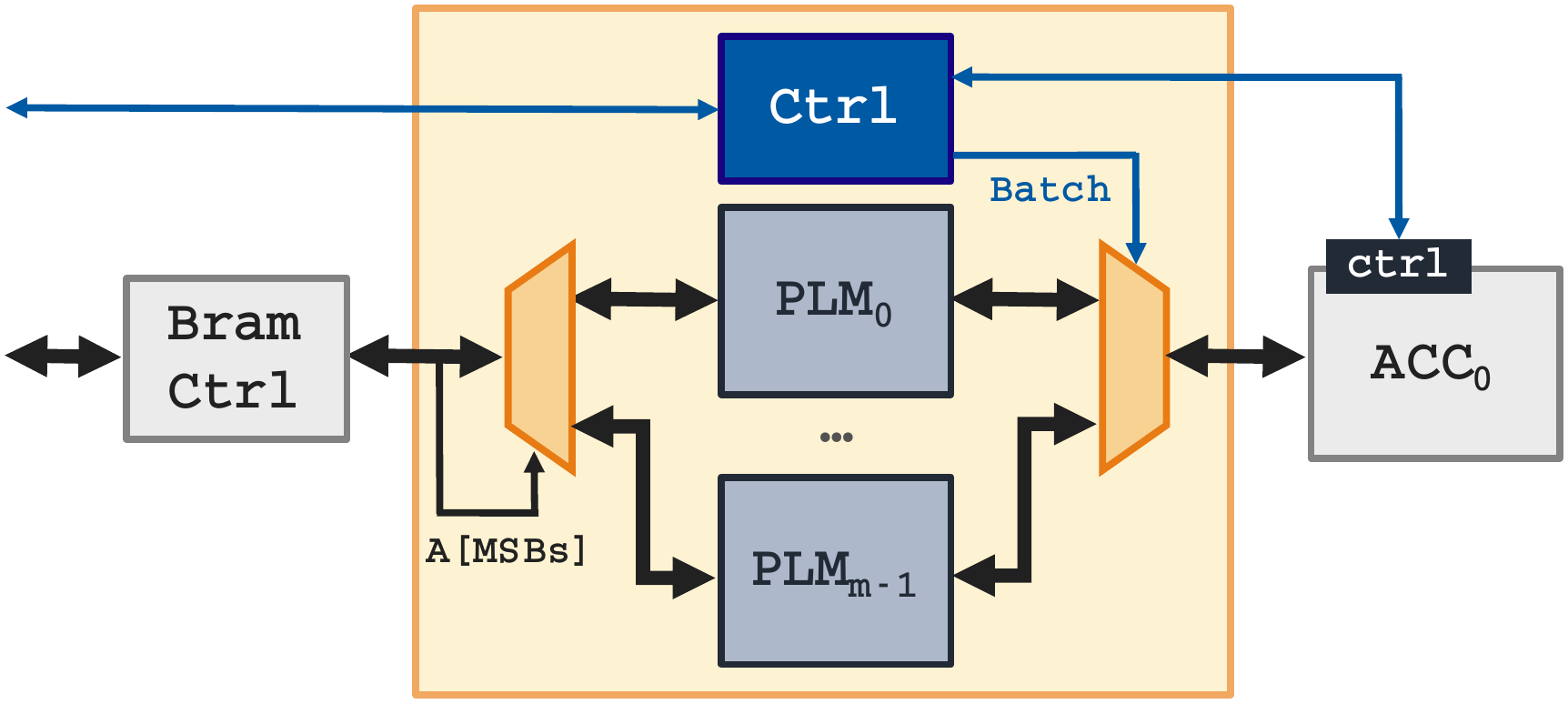}
        }
        \caption{System architecture variations based on $m$ and $k$.}
        \label{fig:sysarchgen}
\end{figure*}

Our compiler-based flow can generate optimized FPGA architectures with several accelerators executing in parallel on different elements (cf.~\Cref{sec:background}). 
To do so, we divide the creation of the target system (cf.~\Cref{sec:target}) in two steps.
In the first step, we generate the accelerator logic (\textit{kernel body}) and the memory subsystem for a single kernel starting from the artifacts generated by the DSL compiler (cf. \Cref{sec:kernel}). 
In the second step, we create a parallel architecture by replicating the memories and the kernels as many times as they can fit into the given FPGA (cf. \Cref{sec:integration}). 
Then, we generate the logic for coordinating the execution and the memory accesses to the different memory instances, along with the corresponding software counterpart for configuring and executing the entire CFD simulation.

\subsection{Kernel Generation}\label{sec:kernel}

To separate the generation of the computational part and the PLM units we export all memory elements from the accelerator. The compiler transforms each memory element (e.g., array or tensor) into an interface parameter of the code to be synthesized.
Figure~\ref{fig:interface} shows an example of resulting C prototype and the corresponding hardware interface generated by HLS.
We implement each array with a PLM unit, i.e., a set of BRAMs that can store the corresponding data (e.g., 121 64-bit elements for array \texttt{S}) and the logic and ports to implement the required read and write accesses. Mnemosyne creates shared PLM units exploiting compiler information (cf. \Cref{sec:mnemosyne}).  

\subsubsection{High-Level Synthesis}\label{sec:hls}

We use commercial HLS tools to generate the RTL code from the C code produced by the compiler. 
When using uni- and multi-dimensional arrays as input parameters, existing HLS tools assume the memory is outside the component. 
They generate a standard memory interface, assuming fixed latency when scheduling memory accesses.
We can apply state-of-the-art HLS optimizations (i.e., loop unrolling and pipelining) since they are independent of the memory interface. 
Array partitioning can be also applied to increase the parallelism, demanding multi-port memories that we manage during memory architecture generation.

\subsubsection{Memory Architecture Generation}\label{sec:mnemosyne}

Each memory element is implemented outside the accelerator on BRAMs.
We apply memory sharing to reduce the BRAM requirements of each kernel. 
To this end, we exploit the information computed during liveness analysis (cf. \Cref{sec:liveness}). 
Mnemosyne uses this information to generate zero-conflict memory architectures while guaranteeing fixed latency of the memory accesses. 
It can also create multi-port, multi-bank architectures based on the requested HLS optimizations.

\subsection{System Generation}\label{sec:integration}

After generating the accelerator and the corresponding optimized memory subsystem, we can compute how many replicas can fit into the given FPGA. 
After reserving FGPA resources for interfaces (e.g., AXI controllers), which can be easily pre-characterized, we can define the set of resources $A$ available for the accelerators and extra routing logic.
We can then estimate the resource requirements of the HLS accelerator ($H$) and its memory ($M$) from the reports. So, our system must respect the following equation: 
\begin{equation}
 [H] \cdot k + [M] \cdot m \leq [A]
\end{equation}
We assume $m \geq k$, since accelerators can only execute in parallel if they each have a memory architecture to work with. To simplify the logic around the accelerators and the PLM units, $m$ must be a power-of-two multiple of $k$. This constraint greatly simplifies the system integration logic.

We developed a tool to read the kernel and memory interfaces, the CDFlang metadata, and the board information to automatically create 1) the accelerator instances, 2) the logic to drive the data from the host to the different PLM units and vice versa, and 3) the system description ready for logic synthesis along with the corresponding software host code. 

Given the number of accelerator ($k$) and memory ($m$) replica, we determine how many times to execute each accelerator (parameter $batch = m / k$) and, in turn, how to connect the modules. 
If $k = m$, the memory ports are simply connected between accelerators and memories of equal index, as shown in \Cref{fig:k2m2}. Each accelerator operates on a single PLM element.
If $k < m$, each kernel operates on $batch$ memories, as shown in \Cref{fig:k1m2}. For instance, if $k=2$ and $m=4$, we have that $batch=2$ and each accelerator~(ACC) operates on two PLMs. In the first execution, ACC$_0$ will access PLM$_0$ and ACC$_1$ will access PLM$_2$. On the second execution ACC$_0$ will access PLM$_1$ and ACC$_1$ will access PLM$_3$. This architecture can amortize long setup times for the data transfers.

The CPU host communicates with the accelerator using an AXI-lite memory mapped interface and an interrupt line. Each of the $k$ accelerators use the \texttt{ap\_ctrl} interface which consists of a \texttt{ap\_start} control input, and \texttt{ap\_done}, \texttt{ap\_idle}, \texttt{ap\_ready} status outputs. To be able to control $k$ accelerators using a single AXI-lite interface to the CPU, we implemented an AXI-lite peripheral to receive the AXI transactions and update the memory mapped registers as if the CPU was interacting with a single kernel generated by HLS with an AXI-lite control interface. To start execution, the host writes a \texttt{start} command to a memory mapped register. When all of the $k$ accelerators are ready to begin, the AXI-lite peripheral broadcasts the \texttt{start} signal to all accelerators. Once each of the $k$ accelerators has signaled that it is done processing, the AXI-lite peripheral raises the interrupt line back to the CPU. 
When a round is completed, the \texttt{batch} counter is incremented up to $m / k$. The \texttt{batch} counter is then forwarded to the memory integration logic, as shown in~\Cref{fig:k1m2}.

The CPU host code executes the accelerator for the total number of elements in the CFD simulation ($N_e$), requiring  $N_e / m$ main loop iterations.
This loop includes the input data transfers, execution, and output data transfers. 
The CPU transfers the input array data for $m$ points through the AXI interface. 
$m$ instances of each array are transferred to power-of-two aligned addresses. 
Then, in a loop which executes $m/k$ times, the start command is sent over the AXI-lite control interface, triggering the execution of $k$ accelerators. 
The CPU waits for the done interrupt. 
After this loop is finished, $m$ points are complete and ready in the output memories. 
The data is transferred in $m$ output arrays available to the CPU. 

\section{Evaluation}\label{sec:eval}
\pgfplotsset{select coords between index/.style 2 args={
    x filter/.code={
        \ifnum\coordindex<#1\def\pgfmathresult{}\fi
        \ifnum\coordindex>#2\def\pgfmathresult{}\fi
    }
}}
\pgfplotstableset{%
    every head row/.style={before row=\toprule,after row=\midrule},
    every last row/.style ={after row=\bottomrule},
    %every even row/.style={before row={\rowcolor[gray]{0.9}}},
    %every column/.style={precision=3},
}
\pgfplotstableread[col sep=comma,header=has colnames]{figures/area.csv}\areatable
\pgfplotstableread[col sep=comma,header=has colnames]{figures/shrsu.csv}\shrsutable
\pgfplotstableread[col sep=comma,header=has colnames]{figures/noshrsu.csv}\noshrsutable
\pgfplotstableread[col sep=comma,header=has colnames]{figures/cpu.csv}\cputable

We implemented a prototype of our DSL-to-FPGA tool flow for the CFD simulation targeting the Xilinx Zynq UltraScale+ MPSoC ZCU106 board. This system allows us to get preliminary feedback on the challenges of FPGA acceleration for such workloads. The board features a quad-core ARM Cortex-A53 and a xczu7evffvc1156-2 FPGA, which has 504K system logic cells (around 230K LUTs and 460K FFs) and 312 block RAMs.
%%%
We use Xilinx Vivado HLS 2019.2 for kernel synthesis and Mnemosyne for the optimization of the accelerator's memory. 
We developed an in-house tool for the generation of the system integration logic, the FPGA system description, and the host code. 
We added hardware timers to measure the execution time of the kernel computation with and without the data transfers. 
%%%
We demonstrate the flow on the Inverse Helmholtz operator with a polynomial degree equal to $p=11$.

\begin{figure}[t]
    \centering
\begin{tikzpicture}
    \begin{semilogxaxis}[
            xlabel={$m=k$},
            ylabel={\footnotesize Speedup vs $m=k=1$},
            y label style={at={(axis description cs:0.05,.5)}},
            log basis x=2,
            ymin=0,
            ymax=18,
            xmax=23,
            log ticks with fixed point,
            legend style={ font=\footnotesize,at={(0.02,0.98)},anchor=north west},
            every tick label/.append style={font=\footnotesize},
            width=1\columnwidth, height=0.25\textwidth,
            cycle list name=mark list,
            ybar,
            bar width=16pt,
            nodes near coords,
            nodes near coords style={font=\footnotesize},
            every node near coord/.append style={
                /pgf/number format/fixed zerofill,
                /pgf/number format/precision=2
            }
        ]
        \addplot[red] table[x=m, y={Kernel su}] {\shrsutable};
        \addlegendentry{Accelerator}
        \addplot[blue,fill=blue!10] table[x=m, y={Total su}] {\shrsutable};
        \addlegendentry{Total}
        %\legend{No Sharing, No Sharing (Theory), Sharing}
    \end{semilogxaxis}
\end{tikzpicture}
    \vspace{-12pt}\caption{Accelerator and total speedup for parallel architectures.}
    \label{fig:shr}
\end{figure}
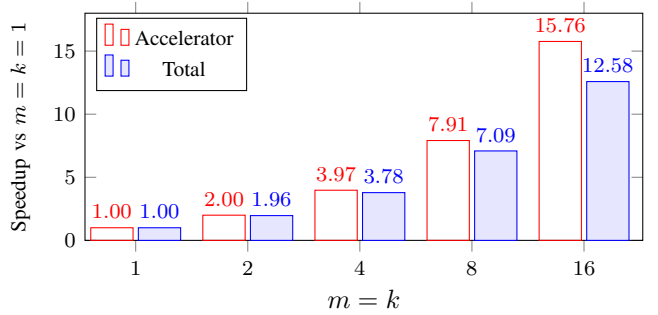

We used CFDlang to generate different C variants having alternative shapes, layout, and compatibility information. 
The CFD accelerator kernel requires around 2,314~LUTs, 2,999~FFs, and 15~DSPs. 
We generated the FPGA systems ignoring sharing compatibilities and using Mnemosyne only as PLM generator. The PLM units for one kernel require 31~BRAMs so we can fit up to $m=8$ units and so $k=8$ kernels. However, when enabling compatibilities obtained from liveness analysis (cf.~\Cref{sec:liveness}), we can fit up to 16 PLM units and kernels (The PLM units for one kernel now require only 18~BRAMs). Resource usage in all cases (from $m=k=1$ to $m=k=16$, when possible) are reported in \Cref{tab:area} (including the rest of the architecture), while \Cref{fig:shr} provides a detailed view on BRAM utilization in all cases. We also generated FPGA systems where the temporary arrays were left inside the HLS accelerator. In these cases, the memory system used 9 BRAMs and the accelerator used 24, for a total of 33 BRAMs, showing that exporting the temporary arrays to allow control over their implementation does allow for better optimization.
All kernels are synthesized at the target frequency of 200~MHz. We executed a prototypical CFD simulation of 50,000 elements with all data in DRAM. 

\begin{table}[t]
    \centering
    \caption{Resource utilization for no memory sharing and memory sharing architectures}
    \label{tab:area}
    \resizebox{\columnwidth}{!}{
    \pgfplotstabletypeset[%string type,
        every head row/.style={
            output empty row,
            before row={
                \toprule
                &$m,k$ & \multicolumn{2}{c}{LUT} & \multicolumn{2}{c}{FF} 
                %& \multicolumn{2}{c}{BRAM} 
                & \multicolumn{2}{c}{DSP}\\
            },
            after row=\midrule,
        },
        every row no 3/.style={
            after row=\midrule,
        },
        columns={shr,m,LUT,LUTper,FF,FFper,
            %BRAM,BRAMper,
            DSP,DSPper},
        columns/shr/.style={string type, column name={}
            %assign cell content/.code={
            %\ifnum\pgfplotstablerow=0
            %    \pgfkeyssetvalue{/pgfplots/table/@cell content}{\multirow{4}{*}{##1}}%
            %\else
            %\ifnum\pgfplotstablerow=4
            %    \pgfkeyssetvalue{/pgfplots/table/@cell content}{\multirow{5}{*}{##1}}%
            %\else
            %    \pgfkeyssetvalue{/pgfplots/table/@cell content}{}%
            %\fi
            %\fi
        %},
        },
        columns/m/.style={column name={}},
        columns/LUT/.style={column type={r@{\hspace{1em}}}},
        columns/LUTper/.style={fixed,zerofill,precision=1,column type={r},%dec sep align,
            postproc cell content/.append style={
                /pgfplots/table/@cell content/.add={(}{\%)},
            }
        },
        columns/FF/.style={column type={r@{\hspace{1em}}}},
        columns/FFper/.style={fixed,zerofill,precision=1,column type={r},%dec sep align,
            postproc cell content/.append style={
                /pgfplots/table/@cell content/.add={(}{\%)},
            }
        },
        %columns/BRAM/.style={column type={r@{\hspace{1em}}}},
        %columns/BRAMper/.style={fixed,zerofill,precision=1,column type={r},%dec sep align,
        %    postproc cell content/.append style={
        %        /pgfplots/table/@cell content/.add={(}{\%)},
        %    }
        %},
        columns/DSP/.style={column type={r@{\hspace{1em}}}},
        columns/DSPper/.style={fixed,zerofill,precision=1,column type={r},%dec sep align,
            postproc cell content/.append style={
                /pgfplots/table/@cell content/.add={(}{\%)},
            }
        },
    ]{\areatable}}
\end{table}

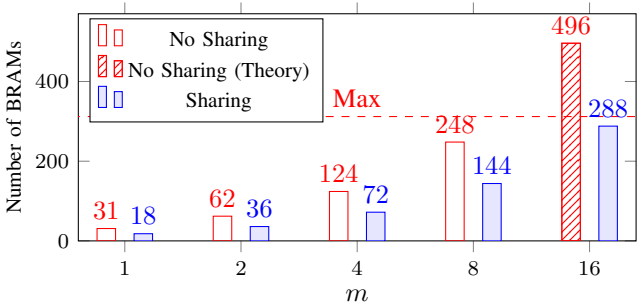
\begin{figure}[t]
    \centering
\begin{tikzpicture}
    \begin{semilogxaxis}[
            xlabel={$m$},
            ylabel={\footnotesize Number of BRAMs},
            y label style={at={(axis description cs:0.05,.5)}},
            log basis x=2,
            ymin=0,
            ymax=570,
            log ticks with fixed point,
            legend style={ font=\footnotesize,at={(0.02,0.98)},anchor=north west},
            every tick label/.append style={font=\footnotesize},
            width=1\columnwidth, height=0.25\textwidth,
            cycle list name=mark list,
            ybar,
            bar width=7pt,
            nodes near coords,
            bar shift auto=0pt,
        ]
        \addplot[red,select coords between index={0}{3}] table[x=m, y=BRAM] {\areatable};
        \addlegendentry{No Sharing}
        \addplot [red,postaction={pattern=north east lines,pattern color=red}, bar shift=-7pt] coordinates { (16,496) };
        \addlegendentry{No Sharing (Theory)}
        \addplot[blue,fill=blue!10,select coords between index={4}{8}] table[x=m, y=BRAM] {\areatable};
        \addlegendentry{Sharing}
        \addplot [red,dashed,line legend,
        sharp plot,update limits=false,
        ] coordinates { (0.1,312) (32,312) }
        node [above] at (2,312) {Max};
        \legend{No Sharing, No Sharing (Theory), Sharing}
    \end{semilogxaxis}
\end{tikzpicture}
    \vspace{-12pt}\caption{BRAM utilization of parallel accelerators w/- and w/o memory sharing.}
    \label{fig:bram}
\end{figure}

First, we tested $k < m$ variants to determine if larger data transfers can reduce communication latency. These experiments did not show much improvements due to limitations in the current implementations of the data transfers. So, we performed all remaining tests with $k=m$. 

\Cref{fig:shr} shows the speed up that we achieved with our parallel architectures. Since memory sharing is transparent to accelerator execution, the values are the same for the two sets of experiments, except in case $m=k=16$ which is possible only with memory sharing. We can see that, as expected, the speedup for accelerator execution is nearly the ideal, $k$. The total speedup is lower due to the communication overhead but can reach up to 12.58$\times$ in case of 16 kernels.

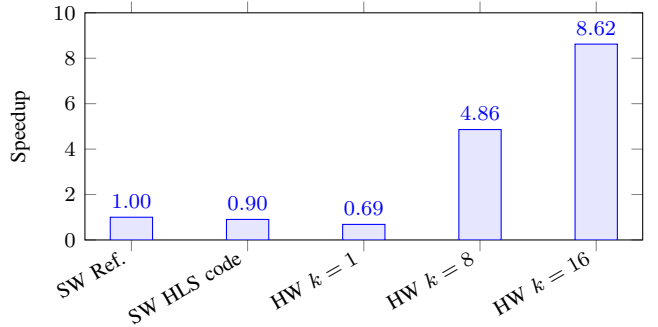
\begin{figure}[t]
    \centering
\begin{tikzpicture}
    \begin{axis}[
            ylabel={\footnotesize Speedup},
            y label style={at={(axis description cs:0.05,.5)}},
            %x label style={at={(axis description cs:0.5,1)}},
            %log basis x=2,
            %xtick={1,2,4,8,16},
            ymin=0,
            ymax=10,
            log ticks with fixed point,
            width=1\columnwidth, height=0.25\textwidth,
            cycle list name=mark list,
            ybar,
            xticklabels={SW Ref., SW HLS code, HW $k=1$, HW $k=8$, HW $k=16$},
            x tick label style={rotate=30,anchor=east,font=\footnotesize},
            every tick label/.append style={font=\footnotesize},
            xtick=data,
            bar width=16pt,
            nodes near coords,
            nodes near coords style={font=\footnotesize},
            every node near coord/.append style={
                /pgf/number format/fixed zerofill,
                /pgf/number format/precision=2
            }
        ]
        \addplot[blue,fill=blue!10] table[x=impl, y={SW su}] {\cputable};
        %\addlegendentry{Total}
        %\legend{No Sharing, No Sharing (Theory), Sharing}
    \end{axis}
\end{tikzpicture}
    \vspace{-12pt}\caption{Speedup compared to software execution on ARM~A53.}
    \label{fig:cpusu}
\end{figure}

For a fair comparison with software execution, we executed a reference implementation of the operator on the ARM~A53 CPU available on the ZCU106 (\textit{SW Ref.} in \Cref{fig:cpusu}). This is the same CPU which performs the data transfers in the hardware execution tests and is configured to run at 1.2~GHz, which is 6$\times$ faster than the kernels running on FPGA.
\Cref{fig:cpusu} shows the comparison for all experiments. The C code given as input to HLS (\textit{SW HLS Code}) is slower on CPU.
We also compared the software with the hardware execution with a variable number of kernels (\textit{HW $k=1$}, \textit{HW $k=8$}, \textit{HW $k=16$}). In case of \textit{HW $k=1$}, the code has 30\% slow-down compared to the software execution because of the faster ARM frequency and the CPU-FPGA data transfers. The best architecture, \textit{HW $k=16$}, executes up to 8.62$\times$ faster than the CPU. 
From the CFD expert viewpoint, all results have been achieved by writing only 9 lines of DSL~(\Cref{fig:helm_op:dsl}) and no particular hardware knowledge (except from board resources).

%\ssol{Bar graph with CPU, no shr k1m1, no shr k8m8, shr k1m1, shr k16m16 speedup vs CPU}

\section{Related Work}
We focused our work on spectral element methods which use tensor expressions. 
Some of existing DSLs such as Tensor Flow Eager \cite{pub:tfe} are  based on software libraries like Tensor Flow or Theano \cite{pub:tf, pub:theano}. These approaches raise the abstraction level, but they offer  limited flexibility in the back-end. DSLs such as Tensor Comprehension (TC) or TVM, on the other hand, were developed to optimize for various platforms. While TC specializes in GPUs, TVM also supports FPGA back-ends. However, this back-end is based on a template architecture and does not offer flexibility for custom tensor expression or for replicating the respective kernels, as also in~\cite{giri2020esp4ml}.

Halide is a representative DSL for stencil operations~\cite{pub:halide}, while Halide-HLS~\cite{pub:h-hls} and HeteroHalide~\cite{pub:hh} offer hardware back-ends. Halide separates computation and scheduling for image processing. This allows applications to be flexibly suited to the target architectures, but also shifts the burden of understanding the platform to the application developer.

% Polyhedral modelling and optimizations.
We leverage the polyhedral model for the scheduling of tensor operators, which was a useful approach in many domains~\cite{poly_lattice_memory,poly-cuda}. We use the polyhedral model to determine access patterns, lifetime ranges, and streaming constraints~\cite{ppcg_scheduling, ppcg_consecutivity}. An automatic flow to generate systolic arrays from the polyhedral model has been proposed in~\cite{autosa}.
Polyhedral models can support the interplay of polyhedral optimizations and hardware back-ends with the memory subsystem. 

%Our Origin
We extended the CFDlang compiler~\cite{teil, rink_gpce18} to target hardware generation.
The IR uses primitive operations without any notion of domain semantics that can be found instead in ML-specific approaches such as TPP~\cite{tpp}.
% MLIR.
Modern frameworks such as MLIR \cite{lattner2021mlir} can ease the combination of independent DSL frontends, tensor middle-ends, and hardware backends. 
Via adapters such as Teckyl~\cite{teckyl}, one can already construct flows that consume TC~\cite{tensor_comprehensions} and process them entirely within MLIR.
The MLIR community is missing a value-based tensor dialect, as opposed to the memory-reference based \texttt{linalg} dialect.
We foresee a similar dialect, possibly based on TeIL.

% DSLs for domain-specific accelerators.
The design of domain-specific accelerators for scientific simulations demands an efficient distributed computing model. Approaches such as \cite{essper_dsl_parallel} use DSLs to explicitly encode inter-kernel pipelining and parallelism on a device. 
When evaluated for CFD codes in~\cite{essper_scalability}, partitioning and communication planning have impact on the scalability of these implementations. 
We focus instead on increasing the utilization and throughput of a single device guided by kernel resource usage.

While HLS simplifies the creation of hardware accelerators~\cite{7368920}, memory optimization, system-level integration, and programmability are still open challenges. 
Several HLS optimizations can improve the use of local memories or physical banks during computation~\cite{10.1145/3061639.3062208,7572091}.
Indeed, PLMs dominate the resource requirements, especially in data-intensive accelerators~\cite{7572091}. 
While independent accelerators can naturally share physical banks, compiler-level analysis can help extracting relevant information about intra-kernel compatibilities.
Many existing FPGA architectures, like IBM CloudFPGA~\cite{7929186} and ESP~\cite{10.1145/3400302.3415753}, separate the computational parts from the interconnection logic. For example, the ESP \textit{services} allow designers to integrate accelerators without any impact to the rest of the system. We extend the same concepts to the local memory architecture. We also enable the creation of parallel architectures. The two approaches are orthogonal.

\section{Conclusions and Future Work}

We presented an end-to-end tool flow to accelerate CFD simulations, combining DSL compiler, commercial HLS and memory optimization tools to seamlessly create a custom accelerator to exploit the intrinsic parallelism of the application. 
On a Xilinx Zynq Ultrascale+ ZCU106, we deployed 16 parallel kernels, achieving a speed-up of 12.58$\times$ over the single-kernel execution and 8.62$\times$ over the ARM CPU (which runs 6$\times$ faster). 
These are promising preliminary results that motivate our future work on more advanced DSL tranformations, MLIR integration, better data transfer strategies, and scaling-up to clusters of larger FPGA boards. 
This is an important step to make FPGA acceleration viable for fluid dynamics.

\begin{comment}
\kf{
Compiler todos in no particular order:
\begin{itemize}
    \item Migrate the IR to MLIR, with a new tensor based (TeIL?) dialect.
    \item Add support for more linear algebra. We really need convolutions if we want to test run-of-the-mill solvers as they are now.
    \item Add support for inferred and parametric tensor shapes.
    \item Find out if the support for interpreted execution with multi-precision types can actually be used to help with accuracy-relevant changes or at least automate testing.
    \item Fix memory layouts so they can be applied before, during \emph{and} after, to make the results more stable.
    \item Make rescheduling results more stable.
    \item Introduce the consecutivity constraints (Tried, blew up, scheduler overwhelmed. Need the cool formulation found in \cite{ppcg_consecutivity}!)
    \item Buffers?
    \item Pragmas! Finally add barvinok to get the heuristic working again.
    \item Also Pragmas! I think there should be a mechanism through which the user, but better yet some pipeline, can pass options to the compiler. Not pragmas necessarily, doesn't have to be in-source, but is maybe easier there.
    \item Come up with at least one strategy per dimension of variance so that sensible or at least stable decisions can be made on variant generation.
\end{itemize}
}
\end{comment}

%
\section*{Acknowledgment}
This project is partially funded by the EU Horizon 2020 Programme under grant agreement No 957269 (EVEREST).
\renewcommand{\bibfont}{\footnotesize}
\printbibliography
\end{document}